\documentclass[useAMS,usenatbib]{mn2e}

\usepackage{graphicx}
\usepackage{times}

\title[The segregation of baryons and dark matter]
  {The segregation of baryons and dark matter during halo assembly}

\author[S. Liao et al.]
{Shihong Liao,$^{1}$\thanks{Email: shliao@nao.cas.cn} Liang Gao,$^{1,2}$ Carlos S. Frenk,$^{2}$ Qi Guo$^{1}$ and Jie Wang$^{1}$ 
\\
$^1$Key Laboratory for Computational Astrophysics, National
Astronomical Observatories, Chinese Academy of Sciences, Beijing,
100012, China\\
$^2$Institute of Computational Cosmology, Department of Physics,
University of Durham, Science Laboratories, South Road, Durham DH1
3LE \\
}
\begin{document}



\maketitle

\label{firstpage}

\begin{abstract}
The standard galaxy formation theory assumes that baryons and dark
matter are initially well-mixed before becoming segregated due to
radiative cooling. We use non-radiative hydrodynamical simulations to
explicitly examine this assumption and find that baryons and dark
matter can also be segregated because of different physics 
obeyed by gas and dark matter during the build-up of the halo. As a
result, baryons in many haloes do not originate from the same
Lagrangian region as the dark matter. When using the fraction of
corresponding dark matter and gas particles in the initial conditions
(the ``paired fraction'') as a proxy of the dark matter
and gas segregation strength of a halo, on average about $25$
percent of the baryonic and dark matter of the final halo are
segregated in the initial conditions. This is at odds with the
assumption of the standard galaxy formation model. A consequence of
this effect is that the baryons and dark matter of the same halo
initially experience different tidal torques and thus their angular
momentum vectors are often misaligned. The degree of the misalignment
is largely preserved during later halo assembly and can be understood
with the tidal torque theory. The result challenges the precision of some semi-analytical approaches which utilize dark matter halo merger trees to infer properties of gas associated to dark matter haloes.
\end{abstract}


\section{Introduction}\label{sec_intro}
The standard galaxy formation theory is based on a two-stage paradigm
put forward by \citet[][]{white1978} and \citet[][]{white1991}: (i)
the dominant mass component, cold dark matter (CDM), collapses by
gravitational instability and forms dark matter haloes hierarchically in the
$\Lambda$CDM cosmological model \citep[see][and references
  therein]{frenk2012}; (ii) baryonic matter (gas) condenses in dark matter
potential wells due to a series of dissipative and nonlinear baryonic
processes (e.g. shock-heating, radiative cooling, etc.), and forms
luminous galaxies; see the reviews of \citet[][]{benson2010} and
\citet[][]{somerville2015}.

In this scenario of galaxy formation, a critical assumption is
that baryons follow dark matter tightly before experiencing radiative
cooling. More specifically, it is assumed that the gas and dark
matter, which later form a virialized halo, are initially well-mixed
and hence distributed in the same Lagrangian region. Under this
assumption, the merger trees of dark matter haloes constructed from
pure dark matter simulations are often used as the skeleton to calculate
baryonic evolution in semi-analytical models \citep[SAs, see
  e.g.][]{kauffmann1999, springel2001, guo2011}. By incorporating
baryonic processes with dark matter halo merger trees, SAs achieve
great successes in explaining a large body of observational
data. We refer the reader to \citet[][]{baugh2006},
\citet[][]{benson2010}, \citet[][]{somerville2015} and
\citet[][]{knebe2015} for general reviews and lists of references of
SAs.

This assumption is fundamental to galaxy formation theory and is not
often questioned. However, given that the underlying physics of gas
and dark matter are not entirely the same, i.e., the former is
collisional and reaches equilibrium through shocks, while the latter
is collisionless and becomes virialized via violent relaxation
\citep[]{lyden1967}; the validation of this assumption is not obvious
for hierarchical assembled CDM haloes. Indeed, recently some studies
have questioned this assumption. For example, \citet[][]{benitez2013} show
that the gas in a low-mass halo can be efficiently removed by ram
pressure when it crosses a large-scale pancake. This ``cosmic web
stripping'' mechanism illustrates that the dark matter and gas content
of a halo could be initially segregated in the absence of radiative
cooling. The other example is that, using non-radiative N-body/SPH
simulations, \citet[]{bosch2002} found a significant misalignment
between the angular momentum vectors of gas and dark matter in haloes,
with a median misalignment angle  $\theta \approx 27.1\degr$ and with
large scatter. This result is also at odds with the well-mixing
assumption discussed above, and questions the popular disk formation
model \citep[see e.g. ][]{fall1980, mo1998}. The results of
  \citet[][]{bosch2002} have been confirmed by other hydrodynamical
  simulations \citep[e.g. ][]{yoshida2003, chen2003, bosch2003, sharma2005,
    croft2009, hahn2010, bett2010, zjupa2017}.

In the disk formation model, the gas, which ultimately ends up in a
galactic disk due to radiative cooling, is assumed to share the same initial specific
angular momentum as its dark matter halo because of the following
reasons: (i) in the classical tidal torque theory
\citep[][]{hoyle1951,peebles1969,doroshkevich1970,white1984,catelan1996},
a halo acquires its angular momentum by tidal torques from the
surrounding inhomogeneities; (ii) in the linear regime, the gas
and dark matter of a halo are initially well-mixed and thus experience
the same tidal torques and have identical angular momentum
vectors \citep[][]{fall1980}.

While the angular momentum misalignment has been widely known, its
origin is still not yet fully understood \citep{sharma2012,
  prieto2015}. In this paper, we perform non-radiative hydrodynamical
simulations to examine explicitly the fundamental assumption of the
mixing in the standard galaxy formation theory. If it does not hold,
i.e. the dark matter and gas of a halo are initially segregated, then
the tidal torques they experience and thus their angular momentum
vectors may not necessarily be identical; this provides a natural
solution to the angular momentum misalignment puzzle.

The paper is orgranized as follows. In Section \ref{sec_sim}, we
present our numerical simulations. We investigate the gas-dark matter
segregation of halo in Section \ref{sec_mismat} and its causes in Section \ref{sec_cause}. As an
application, we use it to explain the angular momentum misalignment
between gas and dark matter of haloes in Section
\ref{sec_misali}. Section \ref{sec_dis} summarizes and discusses our
results. We present numerical convergence studies in the appendix.

\section{Numerical Simulations} \label{sec_sim}
We use a Tree-PM N-body/SPH code, Gadget-2 \citep[]{springel2005},
to perform a set of non-radiative hydrodynamical simulations. The
fiducial simulation is a run with $256^3$ dark matter and $256^3$ gas
particles ($256^3\times 2$) in a periodic box with a comoving length
$L_\mathrm{box}=10$ $h^{-1}\mathrm{Mpc}$ on a side. The reason
we choose such a small volume is that the box size has negligible
effect on the problem studied in this paper \citep[see
  e.g.][]{chen2003, sharma2005, croft2009, hahn2010, bett2010,
  zjupa2017}. The cosmological parameters adopted in the simulations are
$\Omega_m=0.30, \Omega_b=0.04, \Omega_\Lambda=0.70,\sigma_8=0.9,$ and
$n_s=0.96$. Thus, the mass of the dark matter and gas particles
are $m_\mathrm{dm}=4.3\times 10^{6}$ $h^{-1}\mathrm{M}_\odot$ and
$m_\mathrm{gas}=6.6\times 10^{5}$ $h^{-1}\mathrm{M}_\odot$
respectively. The softening lengths for both dark matter
and gas particles are $\epsilon=1$ $h^{-1}\mathrm{kpc}$ in comoving
units, i.e., about $1/40$ of the interparticle separation.

We use the N-GenIC
code\footnote{http://wwwmpa.mpa-garching.mpg.de/gadget} to generate
the initial condition at redshift $z_\mathrm{ini}=127$ assuming the total
matter distribution follows the linear power spectrum given
by \citet[]{eisenstein1998}. In the initial conditions, it is assumed
that the gas follows the dark matter perfectly in phase-space. To
achieve this, the N-GenIC code adopts the following setup. Firstly,
$N_p=256^3$ ``original'' particles are used to sample the total matter
(including both dark matter and gas) density distribution by perturbing the
positions and velocities of a glass particle distribution
\citep[]{white1996} with the Zel'dovich approximation
\citep[]{zeldovich1970}. Then each ``original'' particle is split into a
dark matter and a gas particle by displacing their positions as

\begin{eqnarray}
\bmath{r}_\mathrm{dm} &=& \bmath{r}_\mathrm{ori} + \frac{1}{2}\frac{\Omega_b}{\Omega_m}\bar{L}\hat{\bmath{r}}, \\ \nonumber
\bmath{r}_\mathrm{gas} &=& \bmath{r}_\mathrm{ori} - \frac{1}{2}\frac{\Omega_m-\Omega_b}{\Omega_m}\bar{L}\hat{\bmath{r}},
\end{eqnarray}
where $\bmath{r}_\mathrm{ori}, \bmath{r}_\mathrm{dm},
\bmath{r}_\mathrm{gas}$ are the position of the ``original'', dark
matter and gas particle, respectively, $\bar{L}$ is the mean
interparticle separation of the ``original'' particle set, i.e. $\bar{L}=L_\mathrm{box}/N_p^{1/3}$, and
$\hat{\bmath{r}}=(1,1,1)$. The velocities of the resulting
dark matter and gas particles are set to be identical to the velocity
of their ``original'' particle, i.e.,

\begin{equation}
\bmath{v}_\mathrm{dm} = \bmath{v}_\mathrm{gas} = \bmath{v}_\mathrm{ori}.
\end{equation}
The masses of a dark matter and a gas particle are
\begin{equation}
m_\mathrm{dm} = \frac{\Omega_m - \Omega_b}{\Omega_m}m_\mathrm{ori}
\end{equation}
and
\begin{equation}
m_\mathrm{gas} = \frac{\Omega_b}{\Omega_m}m_\mathrm{ori}
\end{equation}
respectively. By doing so, a perfectly mixed distribution for both
dark matter and gas are obtained. 

To carry out numerical convergence tests, we perform two additional
simulations with $128^3\times 2$ and $512^3\times 2$ particles
respectively starting from the initial conditions generated with the same
random phases as that of our fiducial $256^3\times 2$ run. These two
simulations are evolved to $z=2$. The resolution convergence studies
are presented in Appendix \ref{ap_res}. In addition, we run another
$256^3\times 2$ simulation with the same simulation setup but starting
from a grid initial condition. We confirm that adopting grid or glass
initial conditions does not affect our conclusions. In the main
text, we only present the results from our fiducial simulation
with a glass setup.

We adopt the Amiga Halo Finder \citep[AHF,][]{knollmann2009} to
identify dark matter haloes from our simulations with a virial
overdensity parameter of  $\Delta_\mathrm{vir} = 200$ measured
  with  respect to the mean density. Only haloes with $\geq 2000$ dark
matter particles and $\geq 2000$ gas particles are considered in this
study (see Appendix \ref{ap_res}). In total, there are $227$ haloes in
our sample, for which the lowest mass is $M_{200}\approx 1.0\times
10^{10}$ $h^{-1}\mathrm{M}_\odot$.

\section{Gas and dark matter segregation}\label{sec_mismat}
\begin{figure*} 
\centering\includegraphics[width=500pt]{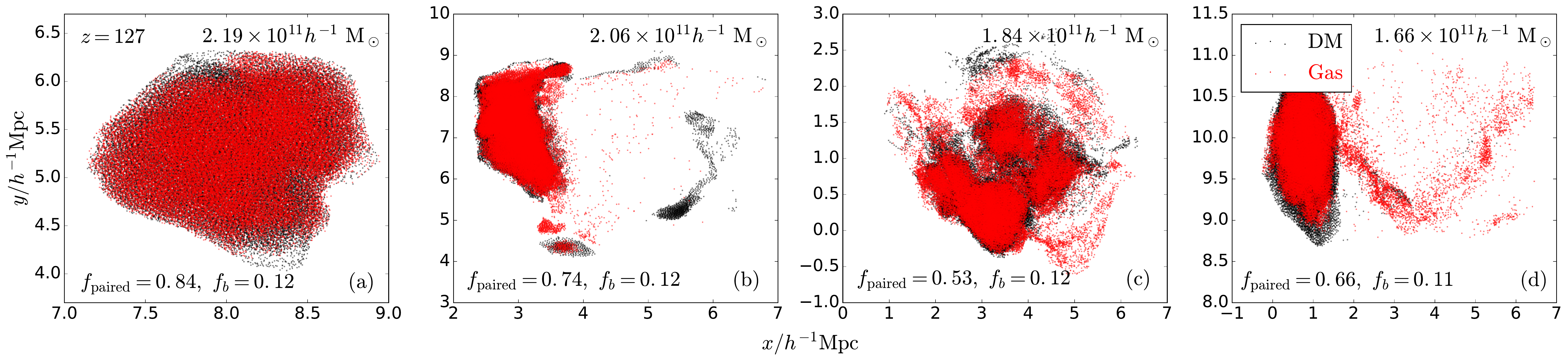} 
\caption{Particle distribution of 4 randomly selected protohaloes in
  the initial conditions $(z=127)$. Dark matter and gas particles are
  shown as black and red dots, respectively. All particles are
  projected on to the $x-y$ plane in comoving coordinates.}\label{protogalaxy}
\end{figure*}

In order to have a direct impression of whether the gas and the dark
matter components of a present day halo are initially segregated, at
$z=0$ we select four random haloes. These haloes have masses of $
\sim 2\times 10^{11}h^{-1}$ $M_\odot$. We then trace all the
particles inside the virial radius, $R_{200}$, of each halo back to
the initial conditions. We show a projection of their positions in
Figure \ref{protogalaxy}. Such a trace-back particle configuration is
dubbed a ``protohalo'' in the rest of the paper. If the dark matter and
gas of a halo are initially well-mixed, the regions occupied by both
components should overlap. Quite surprisingly, as seen in the figure,
the dark matter (black dots) and the gas (red dots) of all selected
haloes are segregated in the initial conditions, although to different
degrees. For the first halo (Panel a), most of the dark matter and
gas particles indeed occupy the same Lagrangian region, but it is
easy to see a lack of gas counterparts on the top and bottom
corners. The second case (Panel b) is quite puzzling. A
disjoint clump of dark matter, a few Mpc away from the dominant
clump, appears in the final halo, whilst the gas particle counterparts
are completely missing. The last two protohaloes (Panels c and d) also
show significant gas-dark matter segregation, with different
strength. Intriguingly, the baryonic mass fraction of each halo,
$f_b\equiv M_\mathrm{gas}/(M_\mathrm{dm}+M_\mathrm{gas})$, as labeled
in each panel, is very close to the universal value,
$f_{b,\mathrm{uni}}\equiv\Omega_b/\Omega_m=0.13$ \citep[see
also][]{crain2007}.

It is reasonable to quantify the gas-dark matter segregation strength
of a halo using a proxy: the particle paired fraction,
$f_\mathrm{paired}$, defined as follows. As described in Section
\ref{sec_sim}, the simulated gas and dark matter particles are
initially split from an ``original'' particle. We define a dark
matter and a gas particle split from the same ``original'' particle
as a \textit{pair}. For each halo at the present day, we count dark
matter-gas pairs, $N_\mathrm{pairs}$, and define a paired fraction
\begin{equation}
f_\mathrm{paired} \equiv \frac{2N_\mathrm{pairs}}{N_\mathrm{tot}},
\end{equation} 
where $N_\mathrm{tot}$ is the total number of particles in a
halo. With such a definition, $f_\mathrm{paired} = 0$ means all gas
and dark matter particles come from different Lagrangian space and so
are completely segregated,  and vice versa for $f_\mathrm{paired} =
1$. The paired fraction values of our four selected haloes
are labeled in Figure \ref{protogalaxy}. Among them, the first
one has the highest value, $0.84$, meaning that $84\%$ of its
particles (dark matter and gas) come from the same Lagrangian space,
while it is only half for the third halo.

\begin{figure} 
\centering\includegraphics[width=240pt]{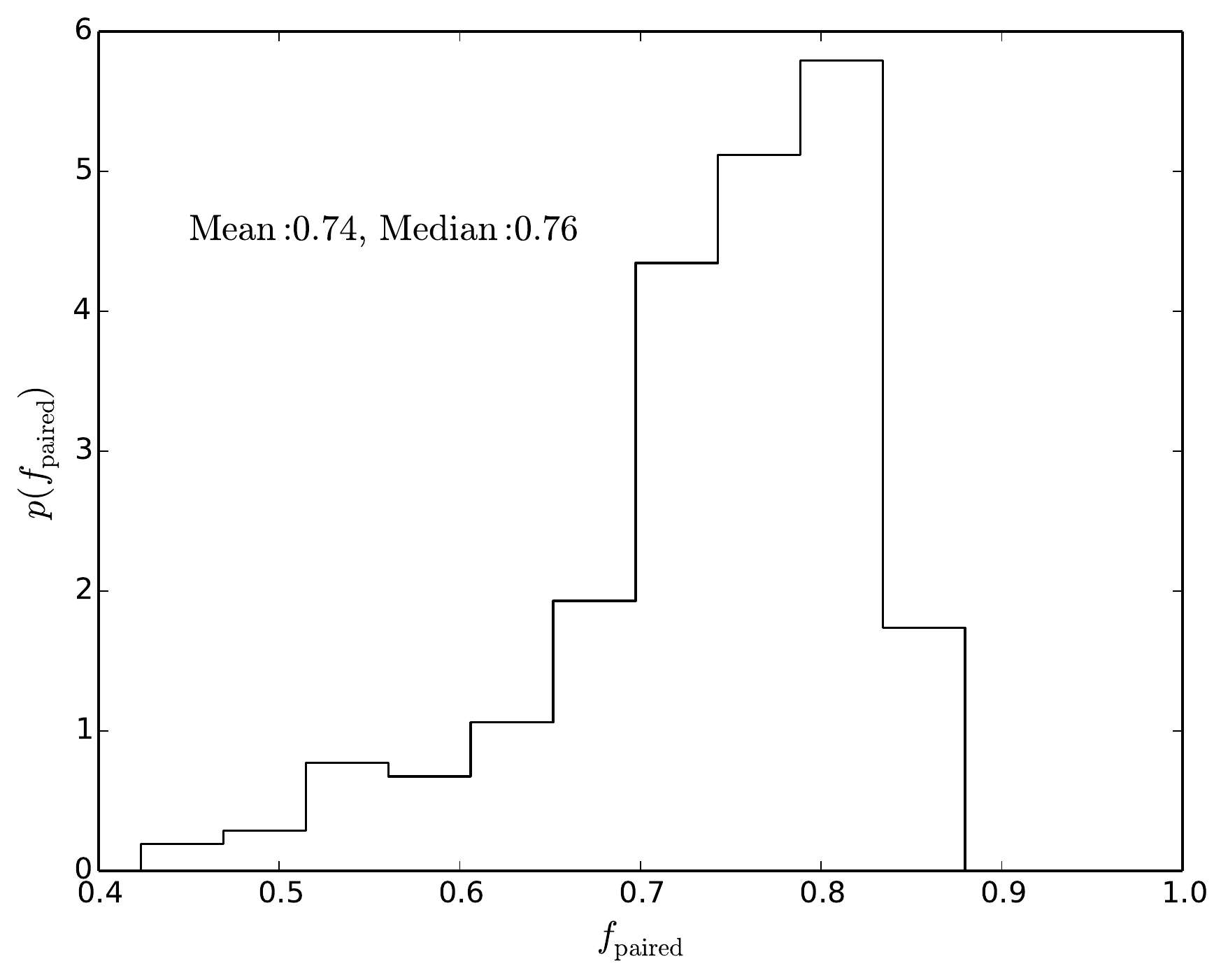} 
\caption{Probability distribution function of $f_\mathrm{paired}$ for
  our full halo sample.}\label{f_pdf}
\end{figure}

The probability distribution function (hereafter PDF) of $f_\mathrm{paired}$ for
our whole halo sample is shown in Figure \ref{f_pdf}. As can be seen,
$f_\mathrm{paired}$ has a fairly broad distribution with a peak
value around $f_\mathrm{paired} \sim 0.8$. The mean and median values
of $f_\mathrm{paired}$ for the whole halo sample are $0.74$ and $0.76$, 
respectively, meaning that, on average $26\%$ of the particles in a
halo are initially segregated. The maximum of $f_\mathrm{paired}$ in
our halo sample is $0.88$, whereas the minimum is as low as $0.42$.

\begin{figure} 
  \centering\includegraphics[width=240pt]{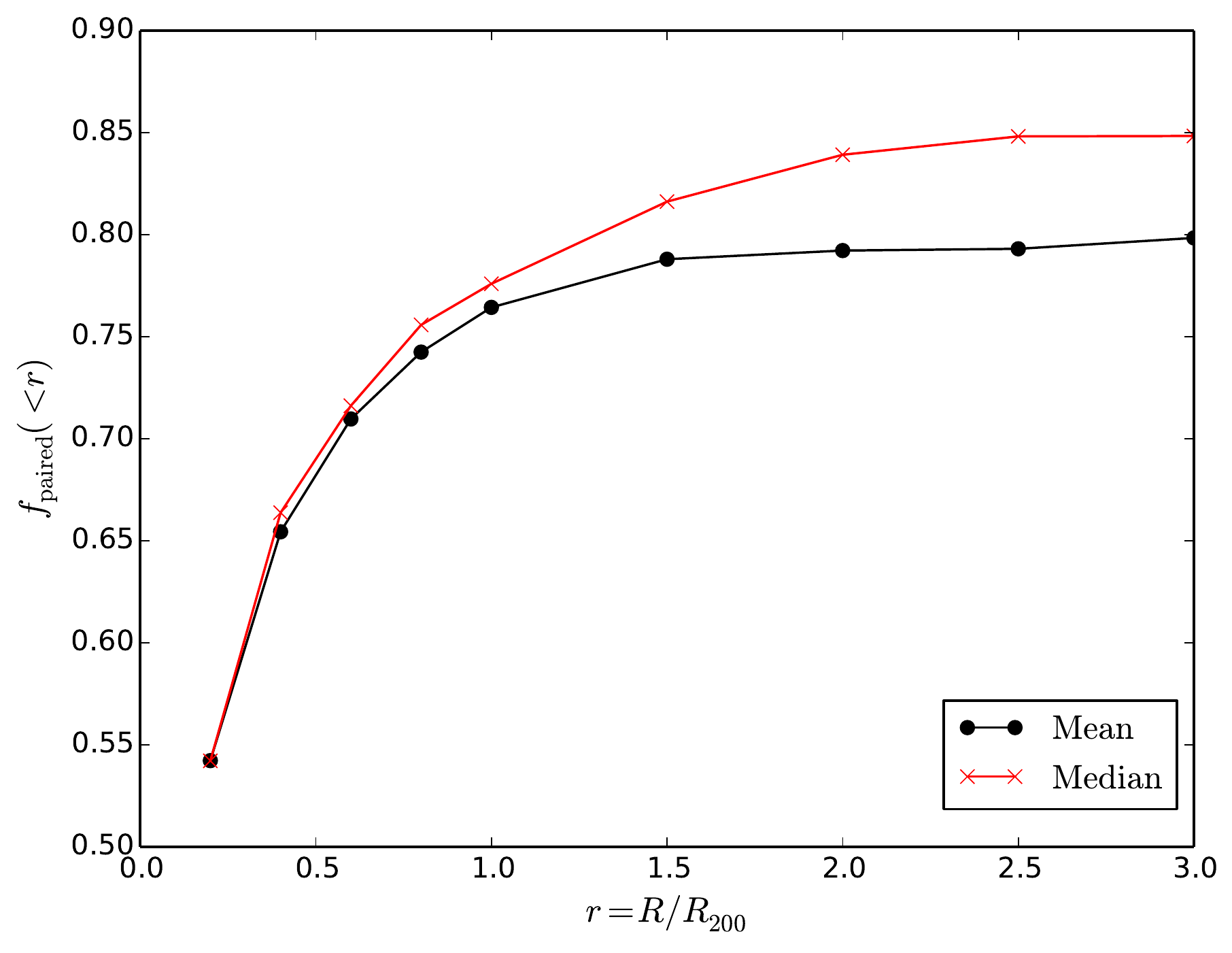}
  \caption{Cumulative paired fraction profiles for the mean (black) and
    the median (red) values of $f_\mathrm{paired}(<r)$ for the 100 most
    massive haloes in our simulation.}\label{f_profile}
\end{figure}

It is interesting to investigate how $f_\mathrm{paired}$ varies with
the distance from the halo centre $R$. Here, the halo centre is defined as the position of the density maximum of the halo. In Figure
\ref{f_profile} we plot the cumulative profile of $f_\mathrm{paired}$ for particles within
different radii from the halo centre for a stacked halo
sample. To ensure that there are enough particles to resolve the
inner halo, we only use the $100$ most massive haloes with
$M_{200}>3.2\times 10^{10}h^{-1}\mathrm{M}_\odot$, which have $\ga
2000$ particles for both dark matter and gas inside $R=0.2R_{200}$, to
compute the profile. Note that, the distance has been
scaled by the virial radius, $R_{200}$. Clearly, the inner particles
tend to be more segregated; on average, half of the particles inside
$R=0.2R_{200}$ lost their {\it partners} in the initial
conditions. Interestingly, even at very large radii, $R=3R_{200}$, 
the unpaired fraction, $1-f_\mathrm{paired}$, is still as large as $\sim 20\%$.

\begin{figure} 
\centering\includegraphics[width=240pt]{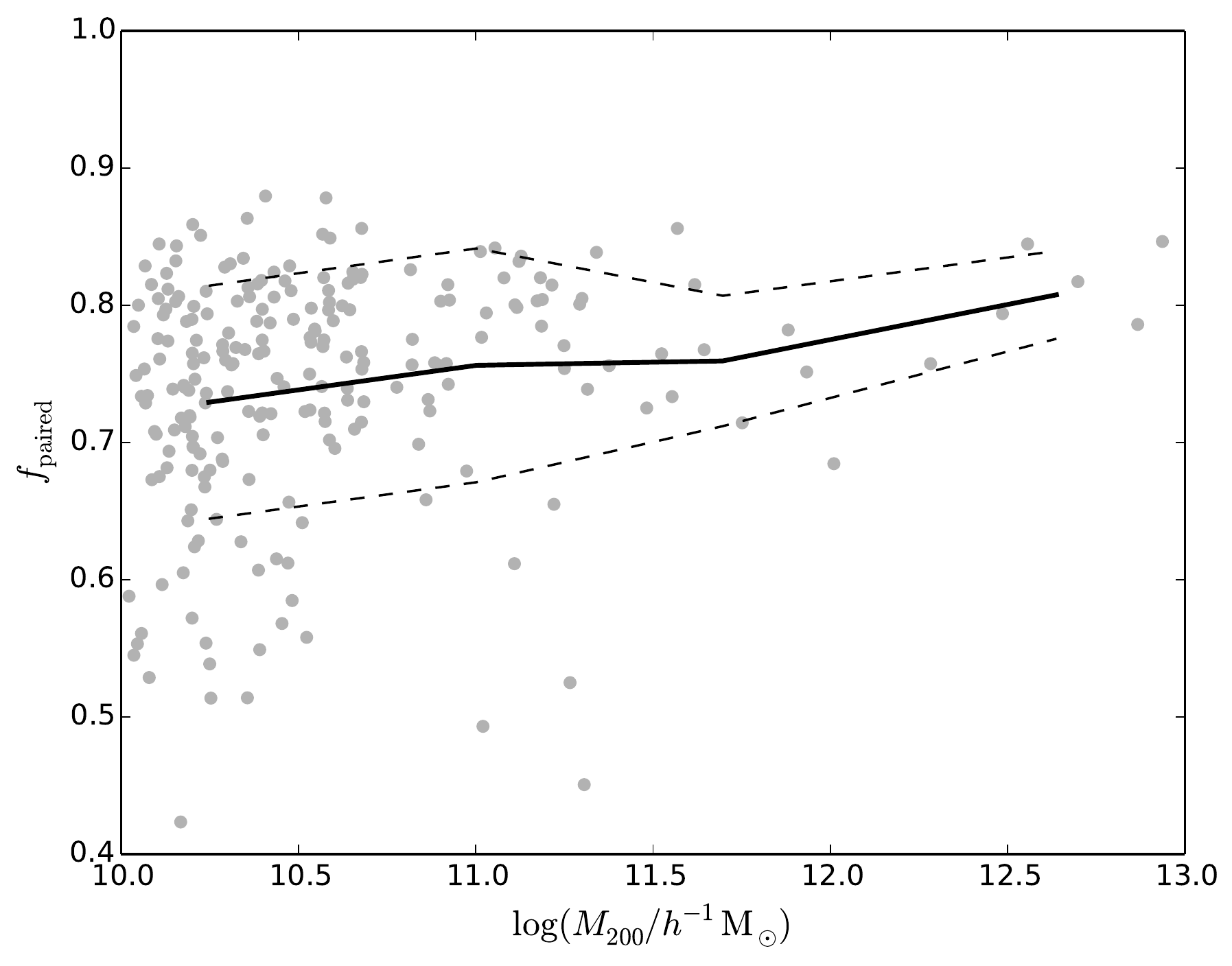} 
\caption{Mass dependence of $f_\mathrm{paired}$. The thick solid line
  (thin dashed lines) shows the mean values (standard deviations) of
  $f_\mathrm{paired}$ in different mass bins. Each grey dot
  represents a halo. }\label{f_mass_dep}
\end{figure}

In order to examine whether $f_\mathrm{paired}$ depends on halo
mass, we plot $f_\mathrm{paired}$ as a function of halo mass,
$M_{200}$, in Figure \ref{f_mass_dep} for our halo sample. There
is a weak mass dependence of the paired fraction with quite large
scatter. On average, galactic haloes have a mean value of
$f_\mathrm{paired} \sim 0.8$, while it is $f_\mathrm{paired} \sim 0.7$
for dwarf-sized haloes. Note that this weak mass dependence is not
a result of numerical resolution effects, as demonstrated in Appendix
\ref{ap_res}. Presumably, this mass dependence could be due to the
fact that more massive haloes have deeper potential wells and thus are
less affected by the surrounding environment.

\section{What causes the gas-dark matter
  segregation?} \label{sec_cause}

What causes the segregation between gas and dark matter during
  halo assembly? We take a closer look at the assembly history of
  several representative haloes. Our finding is that complex interplay
  between dark matter and gas during nonlinear interactions account
  for it, while the resulting physical processes vary from case
  to case. We list a few major processes below.

\begin{figure*} 
\centering\includegraphics[width=500pt]{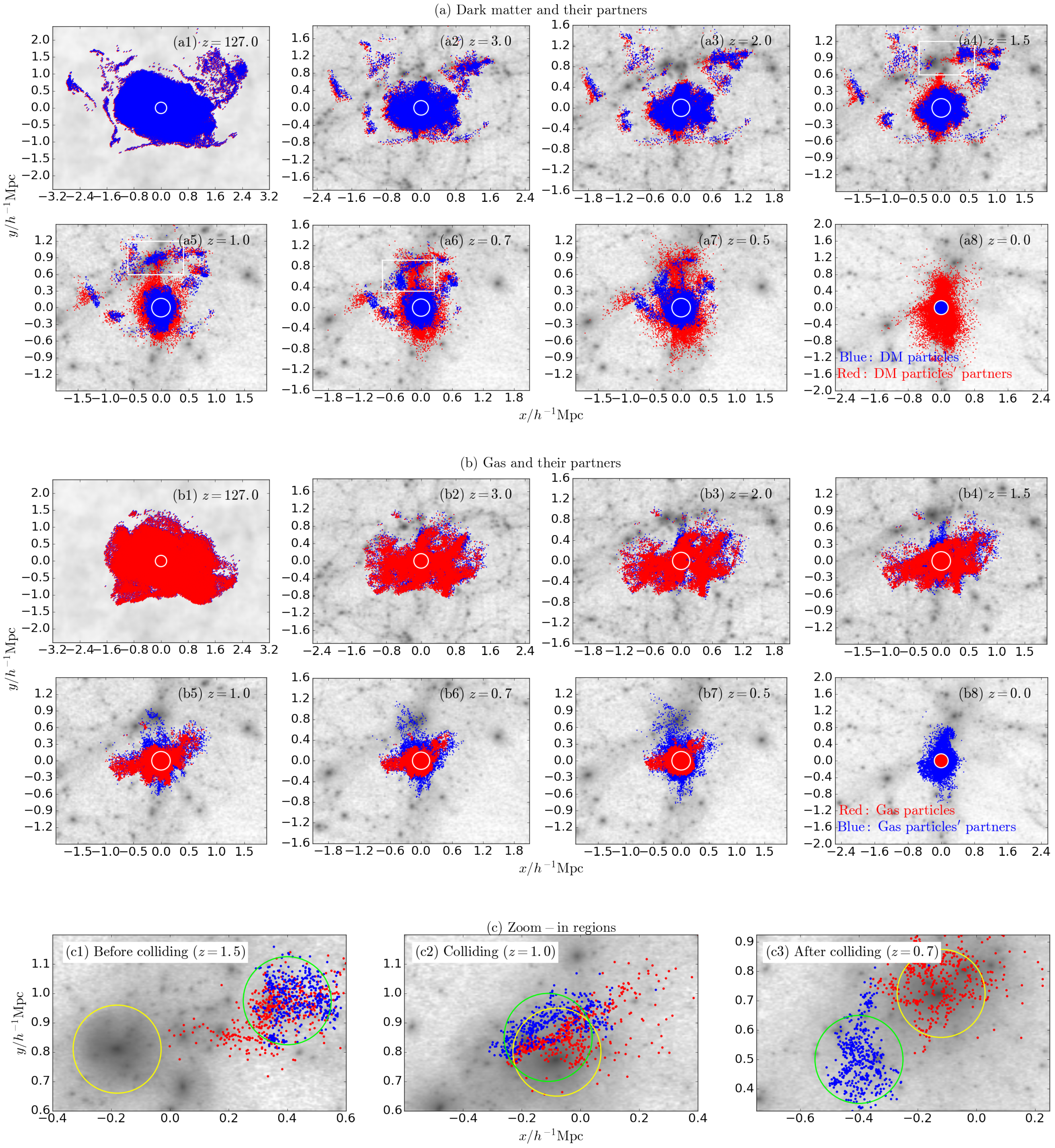} 
\caption{(a) Time evolution of dark matter paritcles (blue dots) of
  halo $\#17$ and their associated gas partners (red dots). The white
  circles represent $R_{200}$ of the final halo. The white rectangles
  in Panles (a4-6) mark the ``zoom-in'' regions which are further
  shown in Panels (c1-3). (b) Similar to Panel (a), but for the gas
  paritcles (red dots) of the same halo and their dark matter partners
  (blue dots). (c) ``Zoom-in'' images of the patches marked in
    Panel (a4), (a5) and (a6) to illustrate the collision of two
    merging haloes (green and yellow circles). Note, to illustrate the
    segregation effect clearly, in Panel (c) we only plot those dark
    matter particles (blue dots) in the halo marked with a green
    circle.}\label{galaxy_vis}
\end{figure*}

In some cases, the segregation is caused by mergers. During the
collision of two merging haloes, the collisional gas particles of one
halo may merge with the other but the collisionless dark matter
counterparts may just pass through and become isolated. The halo may
reaccrete gas from its surroundings. In this case, the halo gas and
dark matter are completely segregated in the initial conditions. As an
example, we present a detailed case for a representative halo, $\#17$,
which has mass of $M_{200}=2.2\times 10^{11}$
$h^{-1}\mathrm{M}_\odot$, radius of $R_{200}=146$
$h^{-1}\mathrm{kpc}$, total particle number of $N_\mathrm{tot}=84307$
and paired fraction of $f_\mathrm{paired}=0.84$, respectively. Figure
\ref{galaxy_vis} illustrates how the gas and dark matter components
from the same Largarigian region get eventually segregated. Panel (a)
shows the temporal evolution of dark matter particles (blue dots)
located within $R_{200}$ of the final $z=0$ halo, as well as their
associated gas partners which are  not necessary inside $R_{200}$ at
$z=0$ (red dots). One can easily see that both components are
initially perfectly mixed, but they start to segregate a little at $z=3$,
and eventually become more and more segregated during the course of
halo clustering. Circles in each plot indicate $R_{200}$ of the final
halo. It is quite striking to see that how extended is the distribution
of gas partners. Panel (b) shows the temporal evolution of
gas particles of the final halo and their dark matter partners. In
Panel (c), we provide ``zoom-in'' images of a patch of Panel (a4), (a5)
and (a6), to illustrate the case of two haloes merging. In this case, the
gas component of two haloes mixed, but the dark matter
counterpart just passed through.

The segregation can be also caused by ``pancake stripping'' as
  pointed out by \citet[][]{benitez2013} and are also observed in our
  own simulation. In this case, when a halo passes through a large-scale
  pancake, its gas component may be entirely stripped by ram-pressure,
  and leave behind a nearly gas-free halo. An additional case occurs
  for haloes located in filaments:  their gas and dark matter
  components initially move along a filament together, but during the
  later evolution, the gas gradually lags behind its dark matter
  counterpart as it experiences additional pressure forces. In this
  case, the dark matter and gas are also disjoint in the initial
  conditions. Other complicated cases also exist, but we do not intend
  to list all of them here. In short, from what we investigated, the
  gas-dark matter segregation is a natural outcome of different
  physics obeyed by gas and dark matter during the non-linear
  evolution.

The gas-dark matter segregation effect discussed above may question
the precision of approaches that use dark matter merger trees to
estimate the evolution of gas residing in dark matter haloes, for
instance, the standard disk formation and semi-analytical galaxy
formation models. As an example, we use this segregation effect to
explain the angular momentum misalignment between gas and dark matter
component of dark matter haloes below.

\section{Misalignment of angular momentum vectors between dark matter
  and gas} \label{sec_misali}

\begin{figure} 
\centering\includegraphics[width=240pt]{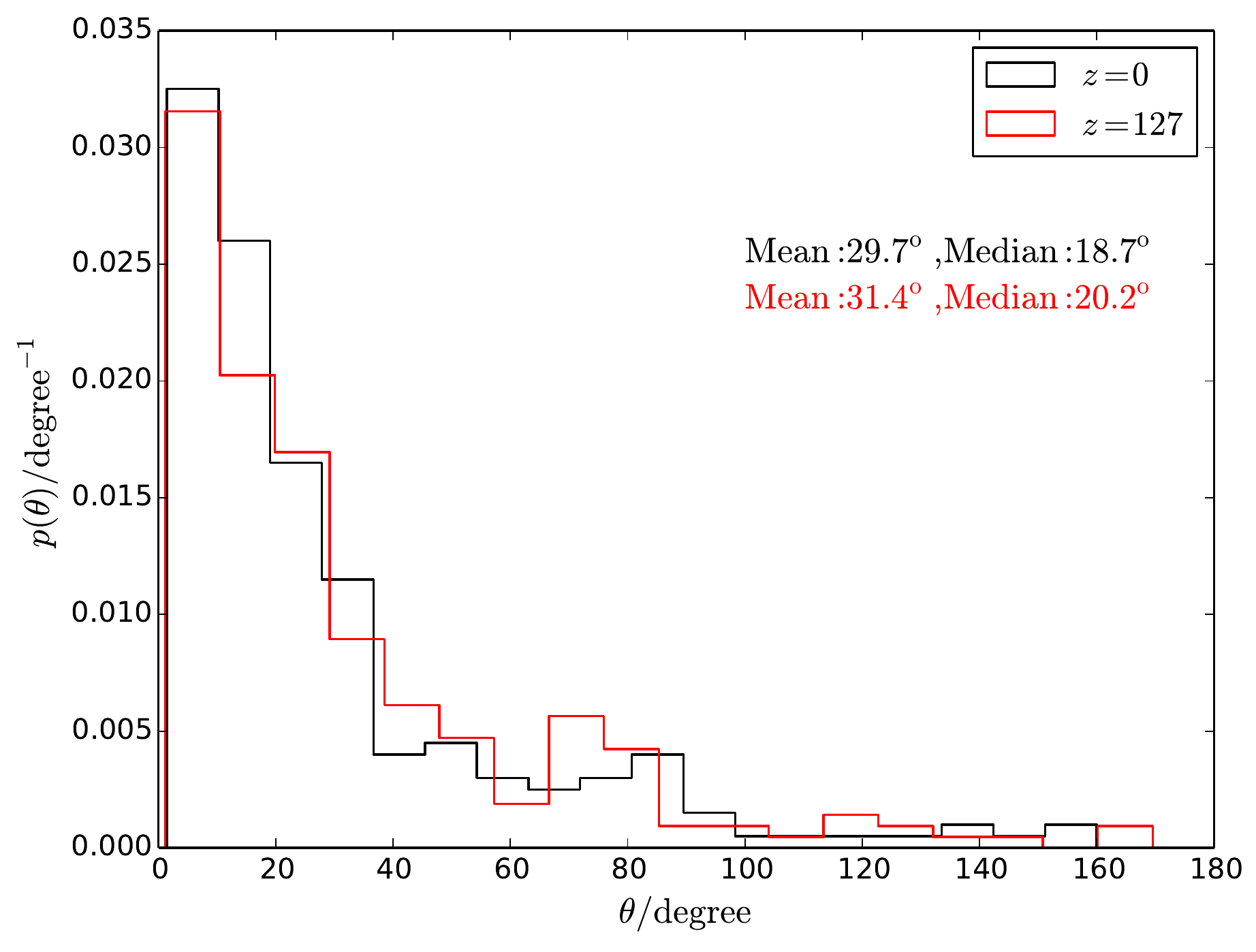} 
\caption{Probability distribution function of the misalignment angle,
  $\theta$, for present day haloes (black) and their protohalo
  counterparts at $z=127$ (red).}\label{theta_pdf}
\end{figure}

As we discussed in previous sections, in the standard galaxy
formation theory, the dark matter and gas components of a halo are
assumed to be perfectly mixed in the initial conditions, and consequently
they are assumed to experience exactly the same tidal torques from
surrounding density fields and so share the same specific angular
momentum. However, as demonstrated in the last section, the gas and
dark matter of a halo are segregated in the initial conditions. To
examine to what extent the segregation effect predicts a
misalignment of the angular momentum vectors between the two
components in the initial conditions, in Figure \ref{theta_pdf} we plot
the PDF of the
misalignment angle, $\theta$,  for the protohaloes of our sample at
$z=127$ as a red solid line. Here $\theta$ is computed as
\begin{equation} 
\theta (z) = \arccos\frac{\bmath{J}_\mathrm{dm}(z)\cdot
  \bmath{J}_\mathrm{gas}(z)}{|\bmath{J}_\mathrm{dm}(z)||\bmath{J}_\mathrm{gas}(z)|},
\end{equation}
where the angular momentum of the dark matter/gas component at
redshift $z$ is
\begin{equation}
\bmath{J}_\mathrm{dm, gas}(z) =
m_\mathrm{dm,gas}\sum_{i=1}^{N_\mathrm{dm,gas}}\left[\bmath{r}_i(z) -
  \bmath{r}_\mathrm{cm}(z) \right] \times \left[\bmath{v}_i(z) -
  \bmath{v}_\mathrm{cm}(z) \right].
\end{equation}
Here, $\bmath{r}_\mathrm{cm}(z)$ and $\bmath{v}_\mathrm{cm}(z)$
are the redshift-dependent center-of-mass position and
  velocity of the particles that are found in a halo at $z=0$,
  respectively.

The misalignment angles $\theta$ of protohaloes have a broad
distribution with a mean (median) value of $31.4 \degr$ $(20.2
\degr)$, which is contrary to expectation from the
well-mixed assumption of gas and dark matter in the initial
conditions. For ease of comparison we also plot the PDF of $\theta$ for
the $z=0$ counterparts as a black solid line. It is quite striking
that the distribution of $\theta$ for the $z=0$ haloes is almost identical
to their counterparts in un-evolved stage in the initial conditions. In
other words, the angular momentum misalignment we see today is already
present in the initial conditions. Note that the PDF of $\theta$ for
our $z=0$ haloes is in good agreement with previous studies \citep[see
e.g.][]{bosch2002, sharma2005}.

\begin{figure} 
\centering\includegraphics[width=240pt]{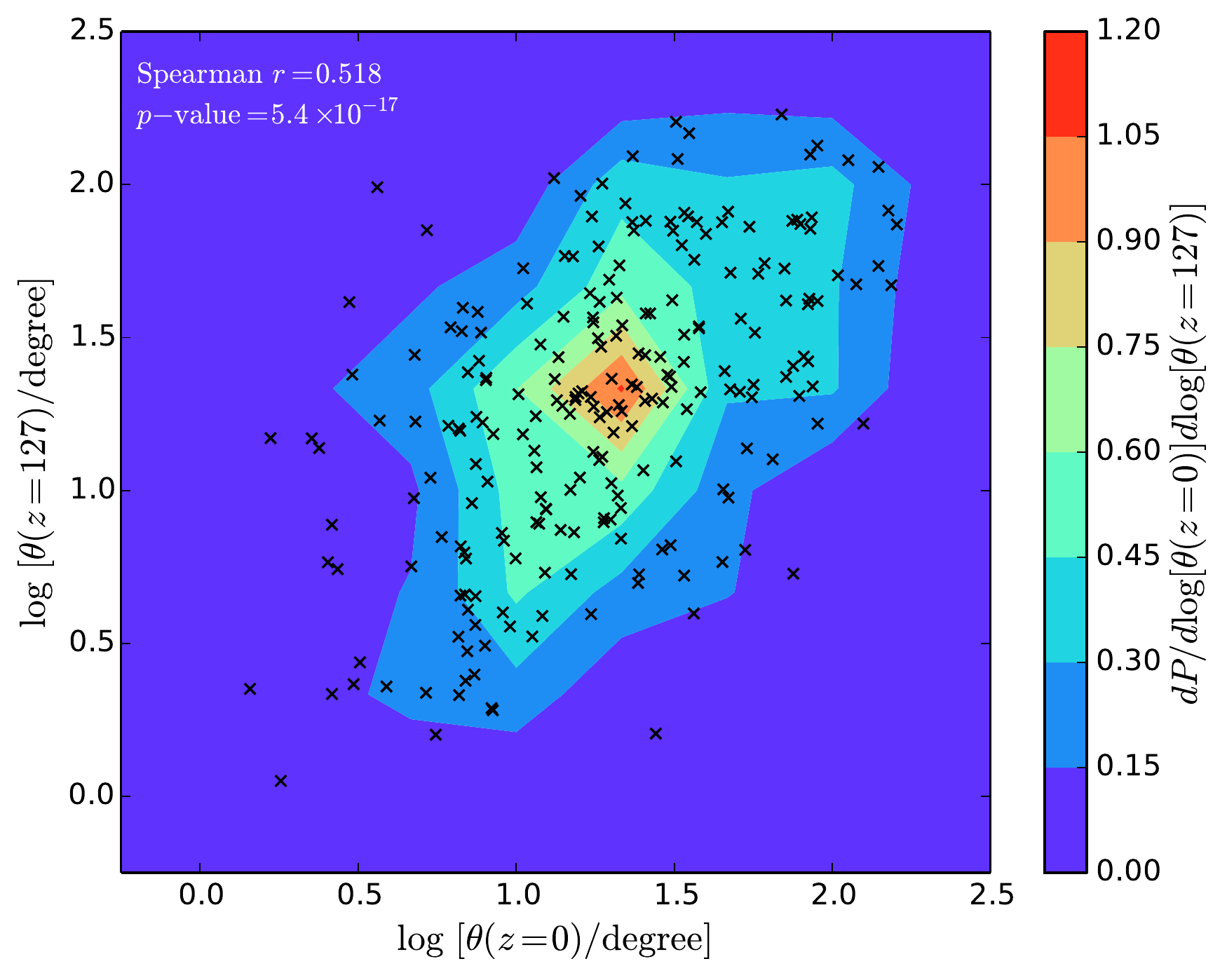} 
\caption{Correlation between $\theta(z=127)$ and $\theta(z=0)$ for
  our halo sample. The contours show the 2D probability distribution
  function calculated from the data points which are marked as
  crosses.}\label{theta_theta_cor}
\end{figure}

\begin{figure}
\centering\includegraphics[width=240pt]{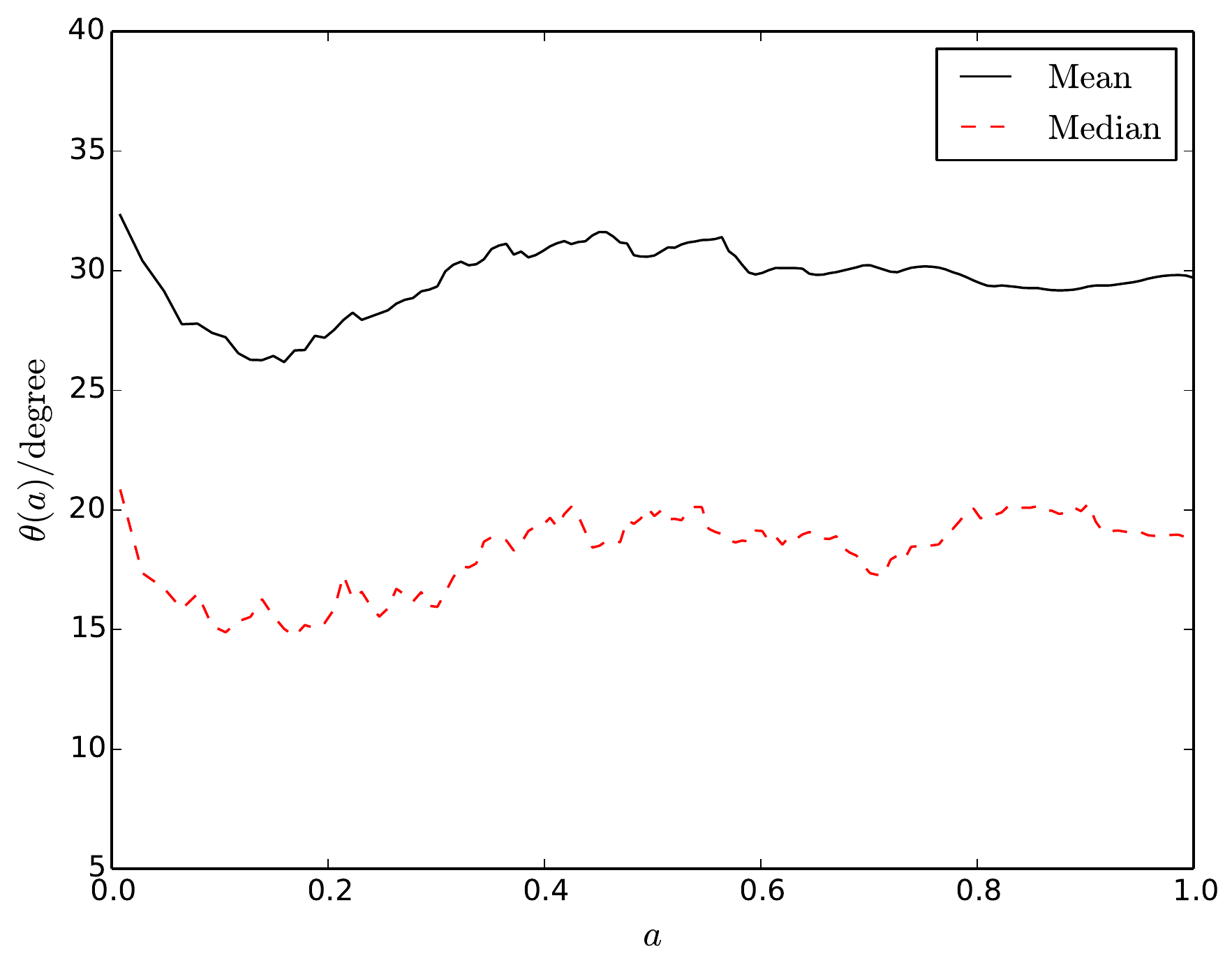} 
\caption{Time evolution of $\theta$ of protohaloes. The mean and
  median of $\theta$ are shown as solid and dashed lines,
  respectively.}\label{theta_evolution}
\end{figure}

In Figure \ref{theta_theta_cor}, we plot the misalignment angle of
each halo in our sample at $z=0$ against its protohalo
counterpart in the initial conditions. Clearly, the misalignment
angles at these two epochs exhibit a strong correlation, with a
Spearman's rank coefficient $r=0.518$ and $p$-value of $5.4\times
10^{-17}$. Note, in order to show the correlation for the data points with small angles more clearly, we plot $\log \theta$ here. But the Spearman's rank coefficient and  $p$-value shown in the upper-left corner are calculated directly from $\theta$. Such a strong correlation  may be understood as follows. According to the tidal torque theory, a halo's angular
momentum is mainly accumulated during the linear evolution and does
not evolve much after collapse because collapsed objects
dramatically reduce their spatial extent and separate from
each other \citep[see e.g.][]{peebles1969, sugerman2000,
  porciani2002}. We thus expect that, once established in the linear
regime, the mean/median misalignment angle of our halo sample will not
vary significantly during the later nonlinear evolution. This is
illustrated in Figure \ref{theta_evolution}, where we present
the time evolution of the misalignment angle between gas and dark
matter of the $100$ most massive haloes in our simulation; the mean and
median values of the misalignment angles are shown at each recorded
snapshot. Clearly, the mean value of the misalignment angle only 
fluctuates mildly with an amplitude smaller than $\sim 5\degr$ during
the whole evolution, consistently with our expectation.

\begin{figure}
\centering\includegraphics[width=240pt]{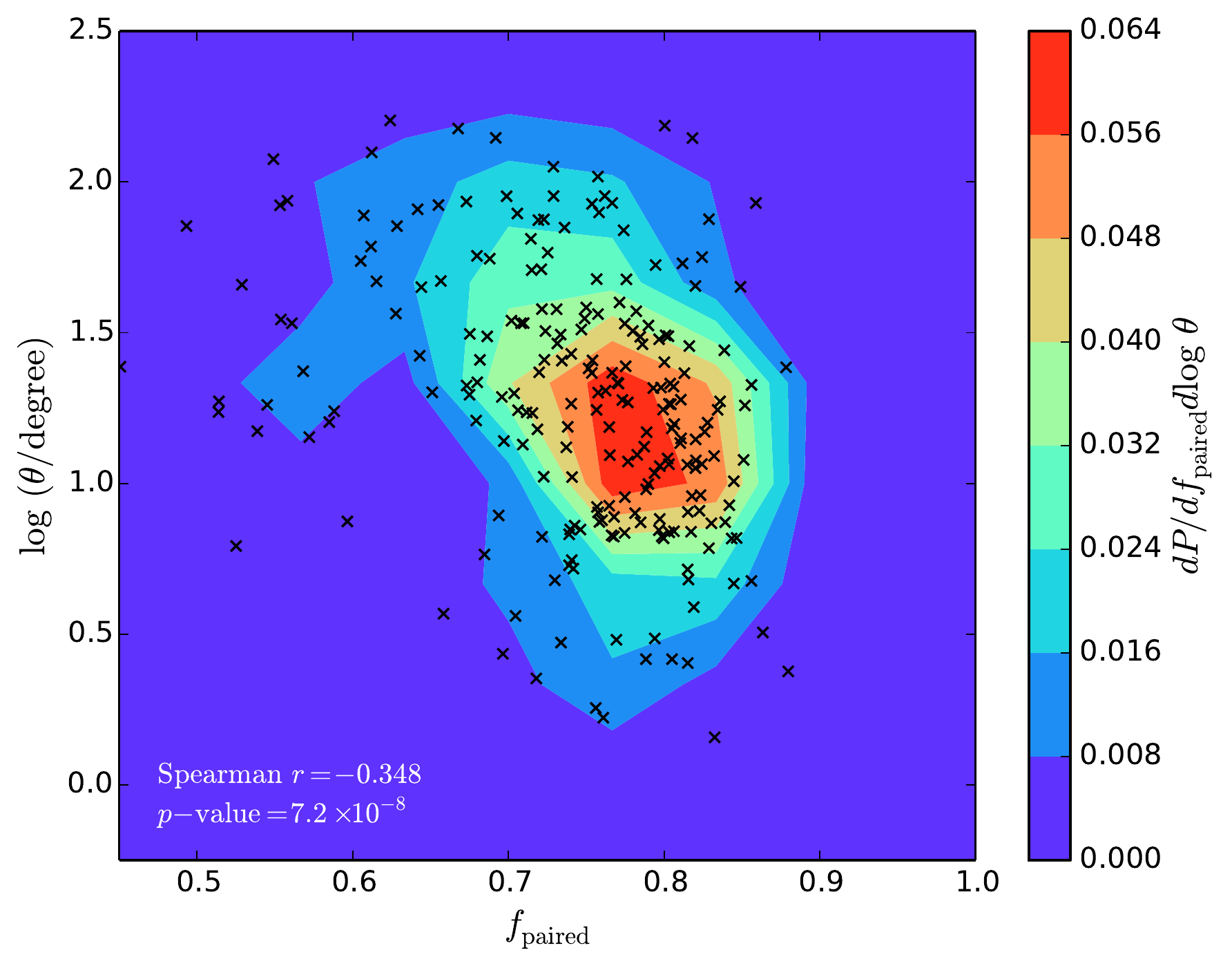} 
\caption{Similar to Figure \ref{theta_theta_cor}, but for the
  correlation between $\theta$ and
  $f_\mathrm{paired}$.}\label{theta_paired_cor}
\end{figure}

To further investigate explicitly the relationship between the
misalignment and the segregation strength, in Figure
\ref{theta_paired_cor} we plot the correlation between the
misalignment angle, $\theta$, and the segregation strength proxy,
$f_\mathrm{paired}$. As shown in the plot, a halo that has a stronger
segregation tends to have a larger misalignment angle. This
correlation is quite strong with a Spearman's rank coefficient
$r=-0.348$ and $p-$value of $7.2\times 10^{-8}$.

\begin{figure} 
\centering\includegraphics[width=240pt]{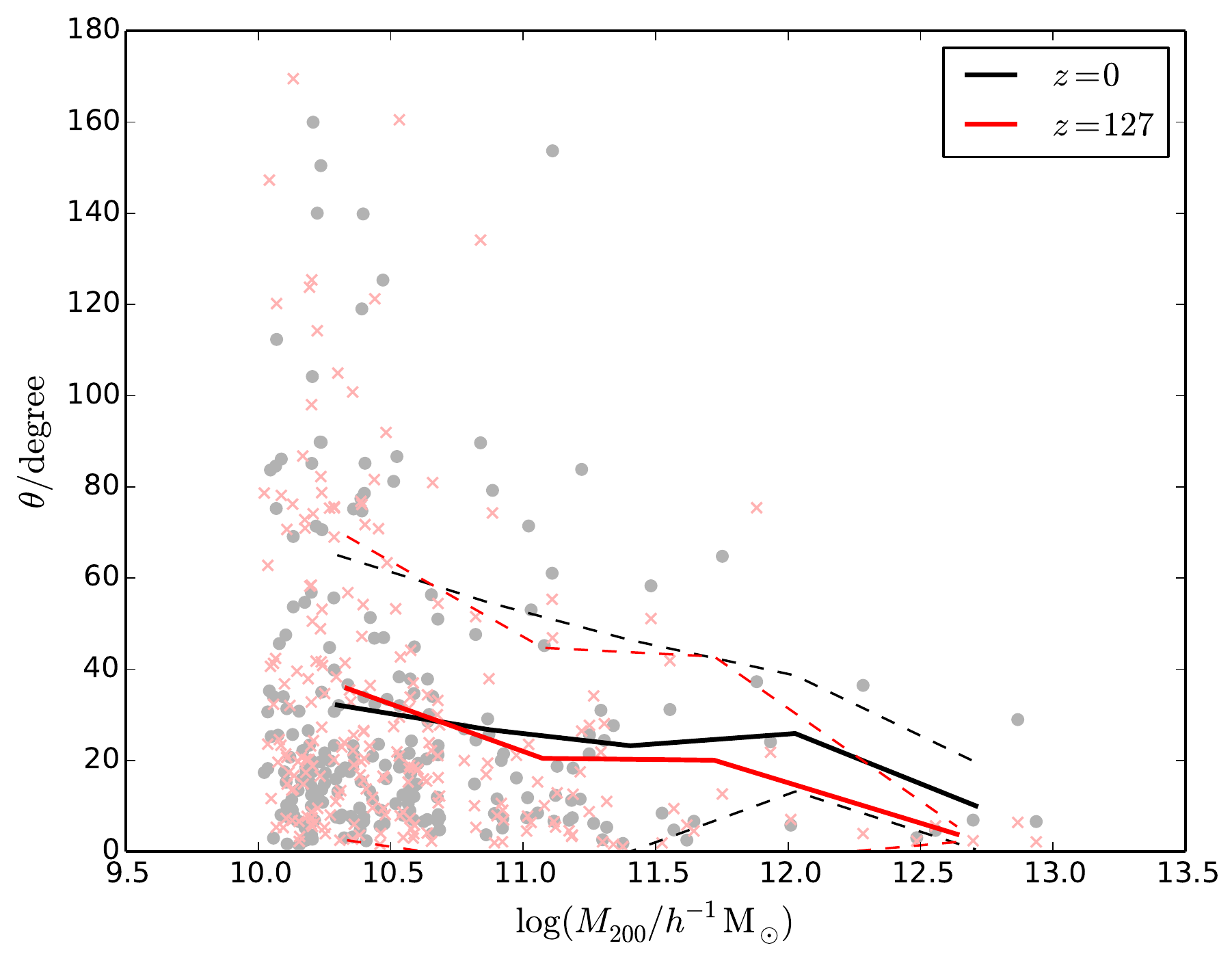} 
\caption{Similar to Figure \ref{f_mass_dep}, but for the mass
  dependence of the misalignment angle, $\theta$, in the initial
  conditions (red) and at $z=0$ (black).}\label{theta_mass_dep}
\end{figure}

As demonstrated in Figure \ref{f_mass_dep}, the segregation
strength proxy, $f_\mathrm{paired}$, depends weakly on halo mass. It is
natural to expect that the misalignment angle should also depend on
halo mass. Since the paired fraction correlates with halo mass,
and the spin misalignment anti-correlates with the paired fraction, we
expect a anti-correlation between the misalignment angle and halo
mass. This expectation is confirmed by Figure \ref{theta_mass_dep}, in
which we plot $\theta$ versus $M_{200}$ for our halo sample, with the
mean value shown as a black solid line. Note that this result is
consistent with previous studies \citep[see e.g.][]{bosch2002}.

In summary, the above results suggest that, as the dark
  matter and gas components of a present day halo are segregated in
  the initial conditions and so have different angular momentum
  vectors. This difference is preserved during later halo assembly and
  can be understood with the tidal torque theory. This naturally
  explains the angular momentum  misalignment of dark matter haloes
  observed at  $z=0$. This explanation is different from that of
\citet[][]{sharma2012} who suggested that the misalignment comes from
galaxy mergers when the intrinsic spins of progenitors are not aligned
with the orbital angular momentum. It also differs from the
explanation of \citet[][]{prieto2015} who argued that additional
pressure torques in the gas lead to the misalignment. Our results,
which are based a large halo sample from a cosmological simulation,
naturally and self-consistently explain various observed facts in
numerical simulations and thus offer a simple and clear explanation
to the puzzling misalignment problem.

\section{Conclusion}\label{sec_dis}
In the current galaxy formation theory, the dark matter and gas
components of a halo are assumed to be well mixed in the initial
conditions before they are segregated due to radiative cooling. In this
study we have used non-radiative N-body/SPH hydrodynamical
simulations to examine this assumption and investigate the
segregation of gas and dark matter during halo assembly.

By tracing particles of present day haloes to the initial conditions,
we find that the dark matter and gas components of haloes are often
initially segregated to varying degrees. When using the paired
fraction as a proxy to measure the segregation strength of haloes, we
find that on average $\sim 25\%$ of the particles (dark and baryonic) in
a halo originates from different Lagrangian regions. The segregation
strength varies with halo mass, with more massive haloes tending to
be less segregated. The paired fraction is about $80\%$ for haloes
with mass larger than $10^{12.5} h^{-1}\mathrm{M}_{\odot}$ and
decreases to $70\%$ for haloes with mass about
$10^{10}h^{-1}\mathrm{M}_{\odot}$. The segregation strength of a halo
is stronger in the inner halo and persists to very outer parts, $\sim 3
\times R_{200}$. Dark matter and gas follow different underlying
physics and this leads to segregation during hierarchical halo
assembly.

The gas-dark matter segregation has important implications for 
galaxy formation theory. As an example, the segregation explains the
misalignment between the angular momentum vectors of gas and dark
matter seen in previous hydrodynamical simulations of galaxy
formation. As the dark matter and gas components of a present day halo are
segregated in the initial conditions, they experience different tidal
torques and therefore end up with different angular momentum
vectors. This difference is preserved during later halo
assembly. Consequently the PDF of $z=0$ haloes and their counterparts
in the initial conditions are almost identical. For individual haloes,
there is a tight correlation between the misalignment angles at $z=0$ and of its
protohalo in the initial conditions, and the segregation strength
proxy, $f_\mathrm{paired}$, correlates with the misalignment angle quite
strongly. All these facts support our argument about the origin of the 
misalignment between the angular momentum vectors of dark matter and gas
in haloes.

The results presented in this paper challenge the precision of
semi-analytical approaches based on the use of dark matter merger
trees to estimate the evolution of gas resident in dark matter
haloes.

\section*{Acknowledgements}
We thank simulating discussions with Prof. Simon White. LG
acknowledges support from the NSFC grant (Nos 11133003, 11425312), and
a Newton Advanced Fellowship, as well as the hospitality of the
Institute for Computational Cosmology at Durham University. CSF
acknowledges ERC Advanced Grant COSMIWAY. QG acknowledges support from the NSFC grant (Nos. 11573033, 11622325), the ``Recruitment Program of Global Youth Experts'' of China, the
NAOC grant (Y434011V01), and a Newton Advanced Fellowship. JW acknowledges the 973 program grant
2015CB857005 and NSFC grant No. 11373029, 11390372.

\appendix
\section{Numerical convergence tests}\label{ap_res}
We have performed two additional simulations with $128^3\times
2$ and $512^3\times 2$ particles respectively. Their initial
conditions have the same random phases as the fiducial $256^3\times 2$
simulation presented in the main text. These simulations have been
carried out to $z=2$. All haloes with at least $10$ dark matter
particles and $10$ gas particles are identified in these simulations.

\begin{figure*}
\centering\includegraphics[width=500pt]{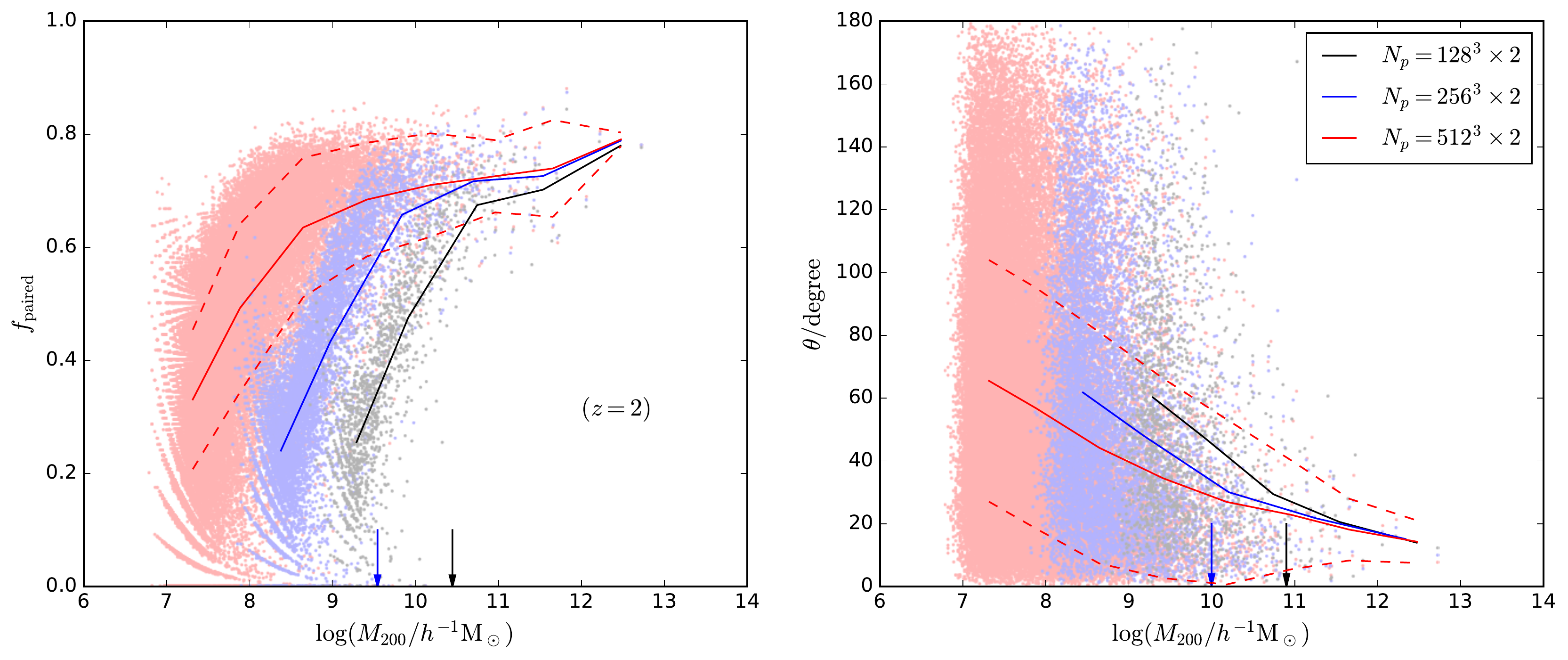} 
\caption{Resolution convergence for haloes' paired fractions (left)
  and misalignment angles (right) at $z=2$. The scatters plot all
  haloes with $\ge 10$ dark matter particles and $\ge 10$ gas
  particles from the $128^3\times 2$ (black), $256^3\times 2$ (blue)
  and $512^3\times 2$ (red) simulations. The solid lines represent the
  mean values at different mass bins. To be clear, we only plot the
  standard deviation for the $512^3\times 2$ simulation (dashed
  lines). The black and blue arrows mark the masses of convergence at
  a $15\%$ level for the $128^3\times 2$ and $256^3\times 2$
  simulations respectively.}\label{resolution_test}
\end{figure*}

In the left panel of Figure \ref{resolution_test}, we plot the paired
fraction, $f_\mathrm{paired}$, against the virial mass, $M_{200}$, of each
halo for the 3 simulations; different colours distinguish different
simulations, as labeled in the legend. The solid lines display the 
respective median values in each mass bin. There is a good convergence
between the results from the $512^3 \times 2$ and $256^3 \times 2$ for
haloes more massive than $M_{200} \ge 3.5\times
10^9$ $h^{-1}\mathrm{M}_\odot$,  and from the $256^3 \times 2$ and
$128^3 \times 2$ for haloes more massive than $M_{200} \ge 2.8\times
10^{10}$ $h^{-1}\mathrm{M}_\odot$.  For both mass scales, haloes with
$\ga 700$ dark matter particles and $\ga 700$ gas particles in the
lower resolution simulation tend to agree with the higher resolution
runs with a difference less than $\sim 15$ percent. The right panel
presents convergence test for the mislighment angles. For haloes
with at least $2000$ dark matter and gas paritlces, their misalighment
angle measuments are free from resolution effects. Hence in this study
we only include dark matter haloes with least $2000$ for both dark
matter and gas paritlces in our halo sample. 

\label{lastpage}


\begin{thebibliography}{101}
\bibitem[Baugh(2006)]{baugh2006} Baugh C. M., 2006, Rep. Prog. Phys., 69, 3101
\bibitem[Benson(2010)]{benson2010} Benson A. J., 2010, Phys. Rep., 495, 33
\bibitem[Ben\'{i}tez-Llambay et al.(2013)]{benitez2013} Ben\'{i}tez-Llambay A., Navarro J. F., Abadi M. G., et al., 2013, ApJ, 763, L41
\bibitem[Bett et al.(2010)]{bett2010} Bett P., Eke V., Frenk C. S., Jenkins A., \& Okamoto T., 2010, MNRAS, 404, 1137
\bibitem[Catelan \& Theuns(1996)]{catelan1996} Catelan P., \& Theuns T., 1996, MNRAS, 282, 455
\bibitem[Chen et al.(2003)]{chen2003} Chen D. N., Jing Y. P., \& Yoshikaw K., 2003, ApJ, 597, 35	
\bibitem[Crain et al.(2007)]{crain2007} Crain R. A., Eke V. R., Frenk C. S., et al., 2007, MNRAS, 377, 41
\bibitem[Croft et al.(2009)]{croft2009} Croft R. A. C., Di Matteo T., Springel V., \& Hernquist L., 2009, MNRAS, 400, 43
\bibitem[Doroshkevich(1970)]{doroshkevich1970} Doroshkevich A. G., 1970, Afz, 6, 581
\bibitem[Eisenstein \& Hu(1998)]{eisenstein1998} Eisenstein D. J., \& Hu W., 1998, ApJ, 496, 605
\bibitem[Fall \& Efstathiou(1980)]{fall1980} Fall S. M., \& Efstathiou G., 1980, MNRAS, 193, 189
\bibitem[Frenk \& White(2012)]{frenk2012} Frenk C. S., \& White S. D. M., 2012, Ann. der Phys., 524, 507
\bibitem[Guo et al.(2011)]{guo2011} Guo Q., White S., Boylan-Kolchin M., et al., 2011, MNRAS, 413, 101
\bibitem[Hahn et al.(2010)]{hahn2010} Hahn O., Teyssier R., \& Carollo C. M., 2010, MNRAS, 405, 274
\bibitem[Hoyle(1951)]{hoyle1951} Hoyle F., 1951, in Problems of Cosmical Aerodynamics, ed. J. M. Burgers \& H. C. van de Hulst (Dayton: Central Air Documents Office), 195
\bibitem[Kauffmann et al.(1999)]{kauffmann1999} Kauffmann G., Colberg J. M., Diaferio A., \& White S. D. M., 1999, MNRAS, 303, 188
\bibitem[Knebe et al.(2015)]{knebe2015} Knebe A., Pearce F. R., Thomas P. A., et al., 2015, MNRAS, 451, 4029
\bibitem[Knollmann \& Knebe(2009)]{knollmann2009} Knollmann S. R., \& Knebe A., 2009, ApJS, 182, 608
\bibitem[Lynden-Bell(1967)]{lyden1967} Lynden-Bell D., 1967, MNRAS, 136, 101
\bibitem[Mo et al.(1998)]{mo1998} Mo H. J., Mao S., \& White S. D. M., 1998, MNRAS, 295, 319
\bibitem[Peebles(1969)]{peebles1969} Peebles P. J. E., 1969, ApJ, 155, 393
\bibitem[Porciani et al.(2002)]{porciani2002} Porciani C., Dekel A., \& Hoffman Y., 2002, MNRAS, 332, 325
\bibitem[Prieto et al.(2015)]{prieto2015} Prieto J., Jimenez R., Haiman Z., \& Gonz\'{a}lez R. E., 2015, MNRAS, 452, 784
\bibitem[Sharma \& Steinmetz(2005)]{sharma2005} Sharma S., \& Steinmetz M., 2005, ApJ, 628, 21
\bibitem[Sharma et al.(2012)]{sharma2012} Sharma S., Steinmetz M., \& Bland-Hawthorn J., 2012, ApJ, 750, 107
\bibitem[Somerville \& Dav\'{e}(2015)]{somerville2015} Somerville R. S., \& Dav\'{e} R., 2015, Annu. Rev. Astron. Astrophys., 53, 51
\bibitem[Springel(2005)]{springel2005} Springel V., 2005, MNRAS, 364, 1105
\bibitem[Springel et al.(2001)]{springel2001} Springel V., White S. D. M., Tormen G., \& Kauffmann G, 2001, MNRAS, 328, 726
\bibitem[Sugerman et al.(2000)]{sugerman2000} Sugerman B., Summers F. J., \& Kamionwski M., 2000, MNRAS, 311, 762
\bibitem[van den Bosch et al.(2002)]{bosch2002} van den Bosch F. C., Abel T., Croft R. A. C., Hernquist L., \& White S. D. M., 2002, ApJ, 576, 21
\bibitem[van den Bosch et al.(2003)]{bosch2003} van den Bosch F. C., Abel T., \& Hernquist L., 2003, MNRAS, 346, 177
\bibitem[White(1984)]{white1984} White S. D. M., 1984, ApJ, 286, 34
\bibitem[White(1996)]{white1996} White S. D. M., 1996, in Schaefer R., Silk J., Spiro M., Zinn-Justin J., eds, Cosmology and Large-Scale Structure, Elsevier, Amsterdam (arXiv: astroph/9410043)
\bibitem[White \& Frenk(1991)]{white1991} White S. D. M., \& Frenk C. S., 1991, ApJ, 379, 52
\bibitem[White \& Rees(1978)]{white1978} White S. D. M., \& Rees M. J., 1978, MNRAS, 183, 341
\bibitem[Yoshida et al.(2003)]{yoshida2003} Yoshida N., Abel T., Hernquist L., \& Sugiyama N., 2003, ApJ, 592, 645
\bibitem[Zel'dovich(1970)]{zeldovich1970} Zel'dovich Ya. B., 1970, A\&A, 5, 84
\bibitem[Zjupa \& Springel(2017)]{zjupa2017} Zjupa J., \& Springel V., 2017, MNRAS, 466, 1625
\end{thebibliography}
\end{document}